\documentclass[10pt]{article}

\usepackage[english]{babel}

\usepackage[dvips]{graphics,color}
\usepackage{newlfont,epsfig}

\usepackage{graphicx}

\begin{document}
\date{}

\begin{center}
{\Large\bf Entanglement reciprocation using three level atoms in a lambda configuration}
\end{center}
\begin{center}
{\large F. C. Louren\c co and A. Vidiella-Barranco \footnote{vidiella@ifi.unicamp.br}}
\end{center}
\begin{center}
{\bf\normalsize{ Instituto de F\'\i sica ``Gleb Wataghin'' - Universidade Estadual de Campinas}}\\
{\bf\normalsize{ 13083-970   Campinas  SP  Brazil}}\\
\end{center}

\begin{abstract}
We propose a scheme in which entanglement can be transferred from atoms (discrete variables) to
entangled states of cavity fields (continuous variables). The cavities play the role of a kind of quantum memory for 
entanglement, in such a way that it is possible to retrieve it back to the atoms. In our method, two 
three level atoms in a lambda configuration, previously entangled, are set to interact with  
single mode cavity fields prepared in coherent states. During the process, one e-bit of entanglement may be
deposited in the cavities in an efficient way. We also show that the stored entanglement may be transferred back
to flying atoms
\end{abstract}

\section{Introduction}

The investigation of quantum entanglement and its manipulation has been of significance given the relevance of 
entanglement as a resource, e.g., for quantum information. A particularly important aspect is the transfer of 
entanglement, specially if one has the purpose of building a quantum network. 
Recently, we have seen the introduction of schemes termed ``entanglement reciprocation",
in which quantum entanglement is transferred from discrete to continuous variables systems and vice-versa 
\cite{lee06,yang06}. Such processes of course widen the possibilities for entanglement storage for further use;
in the proposed methods, ``flying'' quantum systems (atoms) may deposit entanglement to ``stationary'' quantum 
systems (cavity fields). In a similar fashion, entanglement can be subsequently retrieved back to flying qubits (e.g., atoms). 
A peculiar aspect of those schemes is that the cavity field states are actually continuous variable states built from
quasi-classical coherent states, rather than being discrete photon states. Interestingly, in the scheme presented in \cite{lee06}, it is shown the 
possibility of storing one e-bit of entanglement for specific ranges of values of the atom-field interaction time and $\alpha$, the amplitude of the
initial coherent state in the cavities, and having a success probability of 25\%. In fact, atoms interacting with continuous 
variables fields could allow more than one ebit of entanglement to be accumulated, even without using entanglement as a 
resource \cite{palma07}. It would be of importance, though, to seek for more flexible schemes of atom-field and 
field-atom entanglement transfer. A different proposal is found in reference \cite{yang06}, where it is discussed a 
model in which auxiliary classical driving fields are injected inside the cavities (initially in the vacuum state) at 
the same time that they are crossed by the atoms. The authors show that it is possible to achieve a complete 
transfer of entanglement from qubits (atoms) to continuous variables (cavity fields) and vice-versa.
Here we propose an alternative scheme for entanglement reciprocation in which it is possible to deposit one e-bit of 
entanglement in continuous variables states but without the need of classical auxiliary fields. This is made possible
for a wide range of interaction times, and we show that one may also retrieve the stored entanglement with an
efficiency of 100\%. In our scheme, two three-level atoms in a specific lambda configuration are sent across two 
spacially separated cavities previously prepared in single mode coherent states. The atomic upper level is assumed to be
highly detuned from the field, in such a way that we end up with an effective Hamiltonian describing the atom-field
coupling inside the cavities. After the interaction is accomplished and the atoms leave the cavities, they have their internal states measured. 
As a result, the cavities collapse in entangled continuous variables states having up to one e-bit 
of entanglement. Once entanglement is stored in the system of cavities, we show that it can be fully retrieved, for 
instance, by two independent flying atoms crossing the cavities again, i.e., the scheme here proposed allows 
entanglement reciprocation in a straightforward way.  
Our paper is organized as follows: In Sec. II we present our model with its solution. In Sec. III
we describe the process of entanglement transfer from atoms to the field, and in Sec. IV the reverse process. 
We present our conclusions in Sec. V. We also include an Appendix with a detailed derivation of our effective Hamiltonian.

\section{Model}

\begin{figure}[h]
\begin{center}
\resizebox{0.5\columnwidth}{!}{
  \includegraphics{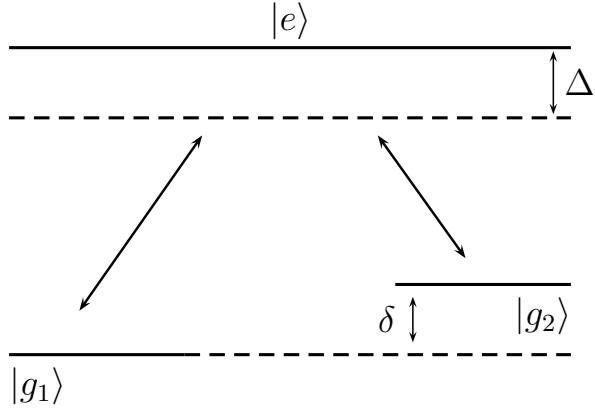}}
  \caption{\label{figure1} Configuration of atomic levels.}
\end{center}
\end{figure}

We consider that initially two identical three-level atoms conveniently prepared in a maximally entangled state
are sent through two independent high-Q cavities. The atoms, in a lambda configuration, are assumed to have two ground states (here 
denoted as $|g_1\rangle$ and $|g_2\rangle$) having an energy separation $E_{g_2} -  E_{g_1} = \hbar\delta$ between them. 
The excited state $|e\rangle$ 
is such that $E_{e} -  E_{g_1} = \hbar\Delta + \hbar\omega$, where $\omega$ is the frequency of the single mode field, as shown in 
figure \ref{figure1}. 
We would like to obtain an effective Hamiltonian in the regime of high atom-field detuning. This may be accomplished if
we take the limit $\Delta \gg g_1,g_2$ ($g_1$, $g_2$ being the atom-field couplings). We also assume two close ground states, in the
sense that $\Delta \gg \delta$. The resulting Hamiltonian (its derivation may be found in the Appendix) 
considering terms up to the first order of $\epsilon = \delta/\Delta$ is
\begin{eqnarray}\label{effhamil}
 H_{e} &=&  H_{e}^{0} +  H_{e}^{1}, \nonumber \\
 H_{e}^{0} &=& \hbar\omega a^{\dag} a 
+ \left[ E_{g_{1}} -\hbar\frac{g_{1}^{2}}{\Delta} a^{\dag} a \right] \sigma_{g_{1}g_{1}}\nonumber \\
&+& \left[ E_{g_{1}} + \hbar\delta - \hbar\frac{g_{2}^{2}}{\Delta} \left( 1+\varepsilon \right) a^{\dag} a \right] \sigma_{g_{2}g_{2}} ,\nonumber \\
 H_{e}^{1} &=& \hbar \lambda a^{\dag} a (\sigma_{g_{2}g_{1}} +\sigma_{g_{1}g_{2}}), 
\end{eqnarray}
where $\lambda = - g_{1}g_{2} \left( 2+\varepsilon \right)/2\Delta$ is the effective coupling constant,
$a_i^\dag$ ($a_i$ ) are the photon creation 
(anihilation) operators for each cavity, and $\sigma_{g_{1}g_{2}},\ \ \sigma_{g_{2}g_{1}}$ are the transition operators for each 
atom. Note the presence of the Stark shifts which arise in the process of adiabatic elimination of the excited level $|e\rangle$. 
We would also like to remark that in the limit of $\delta\rightarrow 0$, i.e., the ground states having the same energy,
we recover the result of the simple degenerate Raman model \cite{knight86}. In order to simplify the calculations, we assume couplings
of the same order of magnitude, more specifically, $g_2\approx g_1/\sqrt{1+\epsilon}$.
Under such conditions, the time evolution of the atom-field states due to Hamiltonian in equation (\ref{effhamil}) may be summarized by
\begin{eqnarray}\label{evolucaog1g2}
e^{-i  H_{e}t/\hbar}|\alpha,g_{1}\rangle = |\xi_{\alpha}^{+}\rangle|g_{1}\rangle - \frac{1}{2} |\alpha_{-}\rangle|g_{2}\rangle \nonumber \\
e^{-i  H_{e}t/\hbar}|\alpha,g_{2}\rangle = |\xi_{\alpha}^{-}\rangle|g_{2}\rangle - \frac{1}{2} |\alpha_{-}\rangle|g_{1}\rangle,
\end{eqnarray}
where the (unnormalized) states above are \footnote{Non relevant global phase factors have been discarded for simplicity.}
\begin{eqnarray}
|\alpha_{-}\rangle &=& |\alpha '\rangle - |\alpha ''\rangle \label{alpha_-} \nonumber \\
|\alpha_{+}\rangle &=& |\alpha '\rangle + |\alpha ''\rangle \nonumber \\
|\chi_{\alpha}^{-}\rangle &=& |\chi_{\alpha '}\rangle  - |\chi_{\alpha ''}\rangle \nonumber \\
|\xi_{\alpha}^{\pm}\rangle &=& \frac{1}{2}|\alpha_{+}\rangle + e^{-|\alpha|^{2}/2} \left( e^{\pm i\delta t/2} -1 \right) |0\rangle  \mp \frac{\delta}{4\lambda_0}|\chi_{\alpha}^{-}\rangle, \nonumber
\end{eqnarray}
with $|\chi_{\alpha}\rangle \equiv \sum_{n=1}^{\infty }e^{-|\alpha|^{2}/2}\frac{\alpha^{n}}{n\sqrt{n!}} |n\rangle$, 
$\lambda_0\equiv g_1^2/\Delta$, $\alpha ' \equiv e^{-i\omega t}\alpha$, and $\alpha ''\equiv e^{i2g_{1}^{2}t/\Delta} \alpha '$.

\section{Atom-field entanglement transfer}

\begin{figure}[h]
\begin{center}
\resizebox{0.6\columnwidth}{!}{
  \includegraphics{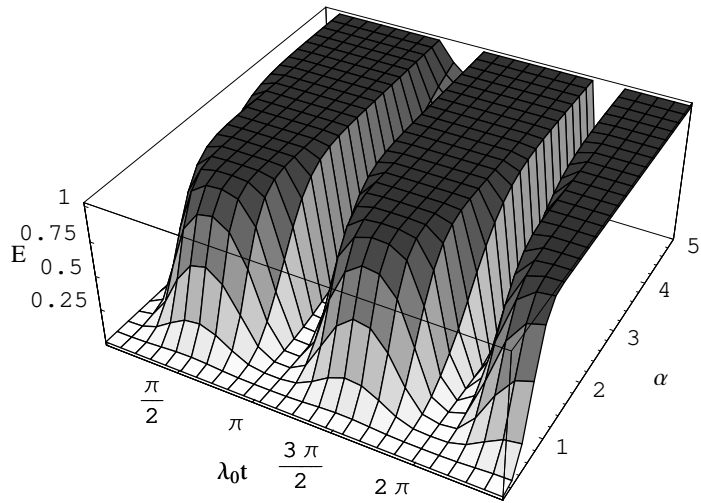}
}\caption{\label{figure2} Entanglement in state $|{C_{21}}\rangle$ as a function of $\lambda_0 t$ and $\alpha$, the initial
coherent amplitude in the cavities. We chose $\delta/\lambda_0 = 1/10$.}
\end{center}
\end{figure}

\begin{figure}[h]
\begin{center}
\resizebox{0.6\columnwidth}{!}{
  \includegraphics{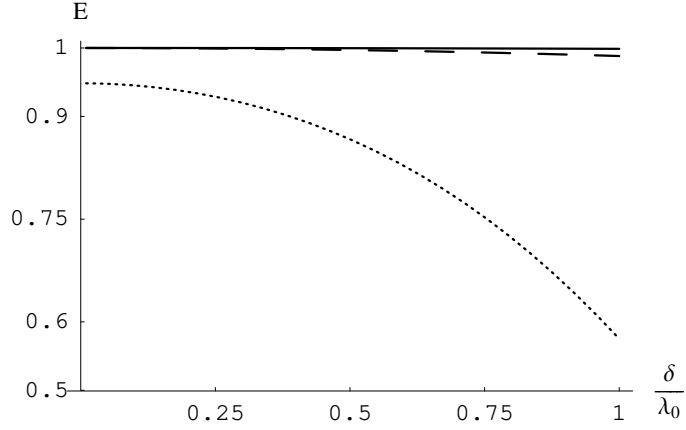}
}\caption{\label{figure4} Entanglement in state $|{C_{21}}\rangle$ at a time $\lambda_0 t = \pi/2$, as a function of 
$\delta/\lambda_0 $. The coherent state amplitudes are $\alpha =1$ (dotted line), $\alpha =3$ (dashed line) e $\alpha =5$ (solid line). 
We note that a significant decrease in entanglement occurs for small values of the initial coherent amplitude, e.g., $\alpha =1$ and for larger values
of $\varepsilon$. However, in such a situation our effective Hamiltonian $ H_{e}$ (equation \ref{effhamil}) is not valid anymore.}
\end{center}
\end{figure}

Now we would like to consider atoms (1) and (2) prepared in a maximally entangled state and the cavities 
$(a)$ and $(b)$ prepared in coherent states $|\alpha\rangle_a$ and $|\beta\rangle_b$, respectively, or
\begin{equation}
|\psi(0) \rangle = |\alpha\rangle_{a}|\alpha\rangle_{b}(|g_{1}\rangle_{1}|g_{2}\rangle_{2} - |g_{2}\rangle_{1}|g_{1}\rangle_{2}).
\end{equation}
After a time $t$ we will have the (unnormalized) state
\begin{eqnarray}
|\psi(t) \rangle &=& -\frac{1}{2}|g_{1}\rangle_{1}|g_{1}\rangle_{2}(|\xi_{\alpha}^{+}\rangle_{a}|\alpha_{-}\rangle_{b} - |\alpha_{-}\rangle_{a}|\xi_{\alpha}^{+}\rangle_{b}) \nonumber \\
&-& \frac{1}{2}|g_{2}\rangle_{1}|g_{2}\rangle_{2}( |\alpha_{-}\rangle_{a}|\xi_{\alpha}^{-}\rangle_{b} - |\xi_{\alpha}^{-}\rangle_{a}|\alpha_{-}\rangle_{b}) \nonumber \\
&+& |g_{2}\rangle_{1}|g_{1}\rangle_{2}( \frac{1}{4} |\alpha_{-}\rangle_{a}|\alpha_{-}\rangle_{b} - |\xi_{\alpha}^{-}\rangle_{a}|\xi_{\alpha}^{+}\rangle_{b}) \nonumber \\
&+& |g_{1}\rangle_{1}|g_{2}\rangle_{2}( |\xi_{\alpha}^{+}\rangle_{a}|\xi_{\alpha}^{-}\rangle_{b} - \frac{1}{4} |\alpha_{-}\rangle_{a}|\alpha_{-}\rangle_{b}). 
\end{eqnarray}
Therefore, after having the atoms measured either in $|g_{1}\rangle|g_{1}\rangle$ or $|g_{2}\rangle|g_{2}\rangle$, the
resulting cavity states will be, respectively
\begin{equation}
|{C_{11}}\rangle = |\xi_{\alpha}^{+}\rangle_{a}|\alpha_{-}\rangle_{b} - |\alpha_{-}\rangle_{a}|\xi_{\alpha}^{+}\rangle_{b},
\end{equation}
or
\begin{equation}
|{C_{22}}\rangle = |\alpha_{-}\rangle_{a}|\xi_{\alpha}^{-}\rangle_{b} - |\xi_{\alpha}^{-}\rangle_{a}|\alpha_{-}\rangle_{b}.
\end{equation}
Interestingly, one may show \cite{wang02} that the states $|{C_{11}}\rangle$ and $|{C_{22}}\rangle$ are maximally
entangled states (contain one ebit of entanglement), irrespectively of the values of the parameters involved. 
This can be clearly seen if the states $|{C_{11}}\rangle$ and $|{C_{22}}\rangle$ were written in an orthogonal basis 
in which they are expressed by a superposition of equally weighted orthogonal states. Nevertheless, in order to obtain 
maximum entanglement, one should avoid specific interaction times, i.e., $2\lambda_0 t \neq 2m\pi,$ with integer $m$, 
since for those times the cavities will be in a product state. Now, if the atoms are measured either in $|g_{2}\rangle_1|g_{1}\rangle_2$ or 
$|g_{1}\rangle_1|g_{2}\rangle_2$, the resulting cavity states will be, respectively 
\begin{equation}
|{C_{21}}\rangle = \frac{1}{4} |\alpha_{-}\rangle_{a}|\alpha_{-}\rangle_{b} - |\xi_{\alpha}^{-}\rangle_{a}|\xi_{\alpha}^{+}\rangle_{b},
\end{equation}
or
\begin{equation}
|{C_{12}}\rangle =  |\xi_{\alpha}^{+}\rangle_{a}|\xi_{\alpha}^{-}\rangle_{b} - \frac{1}{4} |\alpha_{-}\rangle_{a}|\alpha_{-}\rangle_{b}.
\end{equation}
For symmetry reasons one may know that states $|{C_{21}}\rangle$ and $|{C_{12}}\rangle$ have the same amount of entanglement.
Therefore it suffices to calculate the entanglement in one of them, e.g., in $|{C_{21}}\rangle$. In order to do so we first need to
normalize it. From
$$
|\tilde{\alpha}_{-}\rangle = \frac{1}{\sqrt{\langle\alpha_{-}|\alpha_{-}\rangle}} |\alpha_{-}\rangle \quad 
\mbox{and} \quad |\tilde{\xi}_{\alpha}^{\pm}\rangle = \frac{1}{\sqrt{\langle\xi_{\alpha}^{\pm}|\xi_{\alpha}^{\pm}\rangle}} |\xi_{\alpha}^{\pm}\rangle,
$$
we obtain

\begin{eqnarray}
|{C_{21}}\rangle &=& \frac{1}{N_{ C_{21}}} \Big( \frac{\langle\alpha_{-}|\alpha_{-}\rangle}{4} |\tilde{\alpha}_{-}\rangle_{a} |\tilde{\alpha}_{-}\rangle_{b} \nonumber \\
&-&\sqrt{\langle\xi_{\alpha}^{-}|\xi_{\alpha}^{-} \rangle \langle\xi_{\alpha}^{+}|\xi_{\alpha}^{+}\rangle} |\tilde{\xi}_{\alpha}^{-}\rangle_{a} \tilde{\xi}_{\alpha}^{+}\rangle_{b}\Big) ,
\end{eqnarray}
with a normalization factor

\begin{eqnarray}
N_{ C_{21}}&=& \Big[\frac{|\langle\alpha_{-}|\alpha_{-}\rangle|^{2}}{16} + \langle\xi_{\alpha}^{-}|\xi_{\alpha}^{-} \rangle \langle\xi_{\alpha}^{+}|\xi_{\alpha}^{+}\rangle \nonumber \\
&-& \frac{\langle\alpha_{-}|\alpha_{-}\rangle}{4} \sqrt{\langle\xi_{\alpha}^{-}|\xi_{\alpha}^{-} \rangle \langle\xi_{\alpha}^{+}|\xi_{\alpha}^{+}\rangle} ( \langle\xi_{\alpha}^{-}|\alpha_{-} \rangle \langle\xi_{\alpha}^{+}|\alpha_{-} \rangle \nonumber \\
&+& \langle \alpha_{-}|\xi_{\alpha}^{-} \rangle \langle\alpha_{-}|\xi_{\alpha}^{+} \rangle )\Big]^{1/2}.
\end{eqnarray}
Following \cite{wang02}, we may write the eigenvalues  of the reduced density matrix of one cavity, $\rho_{a} = 
Tr_{b}(|{C_{21}}\rangle\langle{C_{21}}|)$, as
\begin{equation}
\mu_{\pm} = \frac{1}{2} \pm \frac{1}{2}\sqrt{1 - 4| adN_{1}N_{2} |^{2}},
\end{equation}
where $a \equiv  \langle\alpha_{-}|\alpha_{-}\rangle/4 N_{ C_{21}} , \ \ d \equiv - \sqrt{\langle\xi_{\alpha}^{-}|\xi_{\alpha}^{-} \rangle \langle\xi_{\alpha}^{+}|\xi_{\alpha}^{+}\rangle}/N_{ C_{21}}$, and $ N_{1} \equiv \sqrt{1 - |
\langle\tilde{\alpha}_{-}|\tilde{\xi}_{\alpha}^{-}\rangle  |^{2}}, \ \ N_{2} \equiv \sqrt{1 - | \langle\tilde{\alpha}_{-}|\tilde{\xi}_{\alpha}^{+}\rangle  |^{2}}$. 
We may then calculate the amount of entanglement in state $|{C_{21}}\rangle$ using the von Neumann entropy, or
\begin{equation}
E = -\mu_{+}\log_{2}(\mu_{+}) -\mu_{-}\log_{2}(\mu_{-}).
\end{equation}
In figure \ref{figure2} we have a plot of the von Neumann entropy $E$ as a function of $\lambda_0 t$ 
and the initial coherent field amplitude $\alpha$. 
We note that the entanglement is strictly zero only at times $\lambda_0t = 0,\pi,2\pi,3\pi,\ldots$, and that
entanglement transfer is relatively robust against variations in the interaction time, especially for large values of $\alpha$, e.g., $\alpha > 2$, as one may see in the plot.
We could ask what would be the dependence of the entanglement in state $|{C_{21}}\rangle$ on the parameter
$\delta/\lambda_0$. Interestingly, as shown in figure \ref{figure4}, specially for larger values of the
amplitude $\alpha$, the entanglement is not very sensitive to variations in $\delta/\lambda_0$.    

\section{Field-atom entanglement transfer}

Now we would like to discuss an example of entanglement transfer from the cavity fields to atoms, a step needed to complete 
the reciprocation process. Let us assume that the cavities had been previously prepared in state $|{C_{11}}\rangle$ 
(for instance, using the method described in the previous section after an interaction time around $t_i=\pi/2\lambda_0$) 
and two independent atoms fly across the cavities, i.e., the atom-field initial (unnormalized) state is
\begin{equation}
|\psi(0) \rangle = |g_{1}\rangle_{1}|g_{1}\rangle_{2}|{C_{11}}\rangle.
\end{equation}
The evolution under Hamiltonian $  H_{e}$ requires the calculation of 
$e^{-i  H_{e}t/\hbar}|g_{1}\rangle|\xi_{\alpha_{1}}^{+}\rangle$, which is given by
$$
e^{-i  H_{e}t/\hbar}|g_{1}\rangle|\xi_{\alpha_{1}}^{+}\rangle = \frac{e^{-i(\omega_{g_{1}} + \delta/2)t}}{2}\left(  |g_{1}\rangle|\mbox{aux-1}\rangle + |g_{2}\rangle|\mbox{aux-2}\rangle \right),
$$
where we have defined the auxiliary states 
\begin{eqnarray}
|\mbox{aux-1}\rangle &\equiv& \Big[|\xi_{\alpha_{1}}^{+}\rangle + |\xi_{-\alpha_{1}}^{+}\rangle +2e^{-|\alpha|^{2}/2} \left( e^{i\delta t/2}-1 \right)|0 \rangle \nonumber \\
&-& \frac{\delta}{4\lambda_0} (|\chi_{\alpha_{1}}^{+}\rangle - |\chi_{-\alpha_{1}}^{+}\rangle)\Big], \nonumber
\end{eqnarray}
and 
$$
|\mbox{aux-2}\rangle \equiv \left[- \frac{1}{2}(|\alpha_{1-}\rangle + |-\alpha_{1-}\rangle) +\frac{\delta}{4\lambda_0} (|\chi_{\alpha_{1}}^{-}\rangle - |\chi_{-\alpha_{1}}^{-}\rangle)  \right].
$$ 
Now, together with equations \ref{evolucaog1g2}, we are able to calculate the joint atom-field state evolution, 
$|\psi(t) \rangle = e^{-i  H_{e}t/\hbar} |\psi(0) \rangle$, which may be written, if we disregard a global phase factor, as 
\begin{eqnarray}
|\psi(t) \rangle &=& |g_{1}\rangle_{1}|g_{1}\rangle_{2}\Big[ |\mbox{aux-1}\rangle_{a} ( |\xi_{\alpha_{1}}^{+}\rangle_{b} \nonumber \\ 
&-& |\xi_{-\alpha_{1}}^{+}\rangle_{b} ) - ( |\xi_{\alpha_{1}}^{+}\rangle_{a} - |\xi_{-\alpha_{1}}^{+}\rangle_{a} )|\mbox{aux-1}\rangle_{b}\Big]  \nonumber \\
&-& \frac{1}{2} |g_{2}\rangle_{1}|g_{2}\rangle_{2} \Big[ |\mbox{aux-2}\rangle_{a} ( |\alpha_{1-}\rangle_{b} - |-\alpha_{1-}\rangle_{b} ) \nonumber \\ 
&-& ( |\alpha_{1-}\rangle_{a} - |-\alpha_{1-}\rangle_{a} )|\mbox{aux-2}\rangle_{b} \Big]  \nonumber \\
&+& |g_{1}\rangle_{1}|g_{2}\rangle_{2} \Big[  -\frac{1}{2} |\mbox{aux-1}\rangle_{a} ( |\alpha_{1-}\rangle_{b} - |-\alpha_{1-}\rangle_{b} ) \nonumber \\
&-& ( |\xi_{\alpha_{1}}^{+}\rangle_{a} - |\xi_{-\alpha_{1}}^{+}\rangle_{a} ) |\mbox{aux-2}\rangle_{b} \Big]  \nonumber \\
&+& |g_{2}\rangle_{1}|g_{1}\rangle_{2} \Big[ |\mbox{aux-2}\rangle_{a} ) ( |\xi_{\alpha_{1}}^{+}\rangle_{b} - |\xi_{-\alpha_{1}}^{+}\rangle_{b} )
\nonumber \\
&+&\frac{1}{2} ( |\alpha_{1-}\rangle_{a} - |-\alpha_{1-}\rangle_{a}  |\mbox{aux-1}\rangle_{b} \Big].
\end{eqnarray} 
Remarkably, provided the cavities are projected to the same coherent states, say 
$|\alpha\rangle_{a}|\alpha\rangle_{b}$, and independently of the interaction time, we may show 
\footnote{This is easily seen as long as the state $|\psi(t) \rangle$ is expressed in terms of states 
$|\alpha_j\rangle$ and $|\xi_j^+\rangle$.} that the terms multiplying
$|g_{1}\rangle_{1}|g_{1}\rangle_{2}$ and $|g_{2}\rangle_{1}|g_{2}\rangle_{2}$ vanish, and the terms multiplying 
$|g_{2}\rangle_{1}|g_{1}\rangle_{2}$ and $|g_{1}\rangle_{1}|g_{2}\rangle_{2}$ differ only by a minus sign. 
Therefore, after the projection of the cavity states, the resulting atomic state will read
\begin{equation}
|\Psi_{-}\rangle = \frac{1}{\sqrt{2}}( |g_{1}\rangle_{1}|g_{2}\rangle_{2} - |g_{2}\rangle_{1}|g_{1}\rangle_{2} ),
\end{equation}
which is a maximally entangled state and is locally equivalent to any other state of the Bell basis. 
Therefore, at least in the ideal case, entanglement reciprocation is possible with 100\% of efficiency, i.e.,
the entanglement of flying atoms previously prepared may be fully transfered to cavities, stored, and then fully retrieved from the cavities back to
a pair of flying atoms. It is worth mentioning that field-atom entanglement transfer is also possible from different initial cavity states, and  
the particular initial state $|{C_{11}}\rangle$ has been chosen for the sake of simplicity. 
We also would like to point out that if the cavities were prepared in one of the states $|C_{12}\rangle$ or $|C_{21}\rangle$, 
the atomic entanglement as a function of the interaction time and the coherent amplitude $\alpha$ has a behaviour similar to the 
one presented in figure \ref{figure2}.  

\section{Conclusions}

We have presented a scheme for entanglement transfer from discrete to continuous variables which would allow, in principle, a complete retrieval of entanglement back to discrete variables. In that scheme, atoms act as flying qubits and entanglement is stored in entangled field states.  
Our model consists of a three-level atom in a lambda configuration interacting with a single quantized mode field, in 
such a way that the lower atomic levels are almost degenerate while the upper level is far off resonance from them. But 
unlike the two-level atom case, that particular three-level configuration has a periodic dynamics, and quantities such 
as the atomic inversion (and entanglement) will show periodicity. In the case of entanglement, we note a recurrence of 
plateaus (see figure \ref{figure2}) indicating maximum entanglement. We remark that due to the 
proximity in energy of the ground levels, we expect similar quantum properties for the generated fields (e.g., 
entanglement), irrespectively of the outcome of the atomic state measurement, as we have demonstrated. 
As a matter of fact, in the special case of exactly degenerate ground states (degenerate Raman model), the generated 
fields will be maximally entangled coherent states (provided the amplitude $\alpha$ is large enough) 
independently of the detected atomic states, as in models of that type the interaction of one atom with a single
cavity naturally allows the generation of superpositions of coherent states.
We may point out some advantages over previous approaches; for instance, there is no need of classical external fields
\cite{yang06} to achieve entanglement reciprocation. Besides, our scheme is very flexible concerning the interaction 
times, i.e., precise interaction times are not required in order to to have full entanglement transfer, 
as seen in figure \ref{figure2}. This is particularly important as it may turn the scheme less sensitive 
to experimental errors and also help to minimize the destructive effects of cavity losses.
Moreover, while the method presented in \cite{lee06} has a success probability of $\sim 25\%$, in our method full 
entanglement transfer will occur irrespectively of the measured states of the outgoing atoms. 
In particular, if we had only considered the cases in which the atoms are detected in 
$|g_{1},g_{1}\rangle$ or $|g_{2},g_{2}\rangle$, then the cavities would get fully entangled (one e-bit) 
independently of both the initial coherent amplitude and the interaction time.

\section{Appendix}

The Hamiltonian describing the coupling of a single mode field with the three-level atom represented in figure \ref{figure1} is
\begin{eqnarray}\label{hamiltoniano_original}
H &=& H_{0} + H_{1}, \nonumber \\
H_{0} &=& \hbar\omega a^{\dag} a +  E_{g_{1}}\sigma_{g_{1}g_{1}} + E_{g_{2}}\sigma_{g_{2}g_{2}} + E_{e}\sigma_{ee}, \\
H_{1} &=& \hbar g_{1} (a \sigma_{eg_{1}} + a^{\dag} \sigma_{g_{1}e}) + \hbar g_{2} (a \sigma_{eg_{2}} + a^{\dag} \sigma_{g_{2}e}), \nonumber
\end{eqnarray}
where $g_{1}$ ($g_{2}$) is the coupling for the transition $|g_{1} \rangle \leftrightarrow |e \rangle$ ($|g_{2} \rangle \leftrightarrow |e \rangle$), $a$ and $a^{\dag}$ are the annihilation and creation photon operators and $\sigma_{jk} = |j \rangle\langle k|$ is the atomic transition operator. We assume that the couplings have the same order of magnitude, $g_{1} \sim g_{2}$. The detunings $\Delta$ and $\delta$, are such that 
$$
\hbar\Delta = E_{e} - E_{g_{1}} - \hbar\omega \ \ \ \ \mbox{and} \ \ \ \ \hbar\delta = E_{g_{2}} - E_{g_{1}}.
$$
The effective Hamiltonian may be obtained from the full Hamiltonian in equation (\ref{hamiltoniano_original}) 
via the the application of a unitary transformation, following well known methods of obtaining
effective Hamiltonians \cite{wu96,moy06}. We may therefore write our transformed Hamiltonian as
$$
H' = e^{S}He^{-S},
$$
where 
$$
S \equiv \frac{g_{1}}{\Delta}(a \sigma_{eg_{1}} - a^{\dag} \sigma_{g_{1}e}) + \frac{g_{2}}{\Delta - \delta }(a \sigma_{eg_{2}} - a^{\dag} \sigma_{g_{2}e}).
$$
We may use the expansion $A' = e^{\xi B}Ae^{-\xi B} = A + \xi[B,A] + \frac{\xi^2}{2!}[B,[B,A]] + \ldots$, to 
rewrite our Hamiltonian as

\begin{eqnarray}\label{expansao_H'}
H' &=& H_{0} + \hbar\left(\frac{g_{1}^{2}}{\Delta} + \frac{g_{2}^{2}}{\Delta - \delta} \right)a a^{\dag}\sigma_{ee} \nonumber \\
&-& \hbar a^{\dag} a \Big[ \frac{g_{1}^{2}}{\Delta}\sigma_{g_{1}g_{1}} + \frac{g_{2}^{2}}{\Delta - \delta}\sigma_{g_{2}g_{2}} \nonumber \\
&+& \frac{g_{1}g_{2}}{2\Delta(\Delta - \delta)}(2\Delta - \delta)( \sigma_{g_{2}g_{1}} +\sigma_{g_{1}g_{2}} ) \Big] 
+ \mathcal{O} \, (g^{2}/\Delta^{2}). \nonumber
\end{eqnarray}
Assuming that $\delta$ is small enough, i.e., $\varepsilon\equiv\delta/\Delta \ll 1$, as well as $g_i/\Delta \ll 1$, so that we can neglect terms of second order in $\varepsilon$ and in $g_i/\Delta$, 
we obtain\footnote{We have taken $\sigma_{ee} = 0$ as state $|e\rangle$ is never populated.} our effective Hamiltonian in equation \ref{effhamil}. 
Note that if $g_{1} = g_{2}$ and $\delta = 0$ the effective Hamiltonian above becomes exactly the degenerate Raman Hamiltonian \cite{knight86}.

The diagonalization of Hamiltonian in equation 
(\ref{effhamil}) gives us the eigenvalues
$$
\mathcal{E}_{+} =\hbar\omega n + E_{g_{1}} + \hbar \delta \left(  \frac{1}{1+\gamma^{2}} \right)
$$
and 
$$
\mathcal{E}_{-} =\hbar\omega n + E_{g_{1}} -\frac{\hbar g_{1}^{2}n}{\Delta} \Big[ 1+\gamma^{2}(1+\varepsilon) \Big]  + \hbar \delta
\left(  \frac{\gamma^{2}}{1+\gamma^{2}} \right),
$$
being $\gamma \equiv g_{2}/g_{1}$. The corresponding eigenstates are
\begin{eqnarray}\label{autoestados}
|n, \varphi_{+} \rangle &=& - c_{1} |n,g_{1} \rangle + c_{2} |n,g_{2} \rangle \nonumber \\
|n, \varphi_{-} \rangle &=& c_{2} |n,g_{1} \rangle + c_{1} |n,g_{2} \rangle, \nonumber
\end{eqnarray}
with 
$$
c_{1} \equiv \frac{\gamma}{\sqrt{1+\gamma^{2}}} \left[ 1+ \frac{\varepsilon}{2(1+\gamma^{2})} - 
\frac{\delta\Delta}{g_{1}^{2}n(1+\gamma^{2})^{2}} \right]
$$
and
$$
c_{2} \equiv \frac{\gamma}{\sqrt{1+\gamma^{2}}} \left[ \frac{1}{\gamma} - \frac{\gamma\varepsilon}{2(1+\gamma^{2})} + \frac{\gamma\delta\Delta}{g_{1}^{2}n(1+\gamma^{2})^{2}} \right].
$$
By applying the condition 
$g_{2} \approx g_{1}/\sqrt{1+\varepsilon}$ to the coefficients $c_1$ and $c_2$ above, we obtain equations \ref{evolucaog1g2}.
\section*{Acknowledgements}
This work was partially supported by CNPq (Conselho Nacional
para o Desenvolvimento Cient\'\i fico e
Tecnol\'ogico), and FAPESP (Funda\c c\~ao de Amparo \`a Pesquisa do
Estado de S\~ao Paulo) grant number 05/57836-7, Brazil.

\end{document}